\documentclass{raa}            

\usepackage{graphicx,times}             

\begin{document}

  \title{The selection of LEGUE disk targets for LAMOST's pilot survey
}

  \volnopage{Vol.0 (200x) No.0, 000--000}      
  \setcounter{page}{1}          

  \author{Li Chen\inst{1}
  \and Jinliang Hou\inst{1}
  \and Jincheng Yu\inst{1}
  \and Chao Liu\inst{2}
  \and Licai Deng\inst{2}
  \and Heidi Jo Newberg\inst{3}
  \and Jeffrey L. Carlin\inst{3}
  \and Fan Yang\inst{2}
  \and Yueyang Zhang\inst{2}
  \and Shiyin Shen\inst{1}
  \and Haotong Zhang\inst{2}
  \and Jianjun Chen\inst{2}
  \and Yuqing Chen\inst{2}
  \and Norbert Christlieb\inst{4}
  \and Zhanwen Han \inst{5}
  \and Hsu-Tai Lee\inst{6}
  \and Xiaowei Liu\inst{7}
  \and Kaike Pan\inst{8}
  \and Jianrong Shi\inst{2}
  \and Hongchi Wang \inst{9}
  \and Zi Zhu\inst{10}
   }

  \institute{Key Laboratory for Research in Galaxies and Cosmology, Shanghai Astronomical Observatory,
  CAS, 80 Nandan Road, Shanghai, 200030, China {\it chenli@shao.ac.cn; houjl@shao.ac.cn}\\
\and
   Key Lab for Optical Astronomy, National Astronomical Observatories, Chinese Academy of Sciences (NAOC)
\and
   Department of Physics, Applied Physics and Astronomy, Rensselaer Polytechnic Institute, 110 8th Street, Troy, NY 12180, USA
\and
   Center for Astronomy, University of Heidelberg, Landessternwarte, K$\ddot{o}$nigstuhl 12, D-69117 Heidelberg, Germany
\and
   Yunnan Astronomical Observatory, Chinese Academy of Sciences, Kunming 650011, China
 \and
 Academia Sinica Institute of Astronomy and Astrophysics, Taipei 10617, China
\and
   Department of Astronomy \& Kavli Institue of Astronomy and Astrophysics, Peking University, Beijing 100875, China
\and
 Apache Point Observatory, PO Box 59, Sunspot, NM 88349, USA.
\and
 Purple Mountain Observatory, Chinese Academy of Sciences, Nanjing 210008, China
\and
   School of Astronomy and Space Science, Nanjing University, Nanjing 210008, China
  }


  \date{Received~~2012 month day; accepted~~2012~~month day}

\abstract{We describe the target selection algorithm for the low latitude disk portion of the LAMOST
Pilot Survey, which aims to test systems in preparation for the LAMOST spectroscopic survey. We use the PPMXL (Roeser et al. 2010)
astrometric catalog, which provides positions, proper motions, B/R/I
magnitudes (mostly) from USNO-B (Monet et al. 2003) and J/H/Ks from The Two Micron All Sky Survey (2MASS, see Skrutskie et al. 2006) as well.
We chose 8 plates along the Galactic plane, in the region $0^\circ<\alpha<67^\circ$ and
$42^\circ<\delta<59^\circ$, that cover 22 known open clusters with a range of ages.
Adjacent plates may have small overlapping. Each plate covers an area $2.5^\circ$ in radius,
with central star (for Shack-Hartmann guider) brighter than $~ 8^{\rm th}$
magnitude. For each plate, we create an input catalog in the magnitude range
$11.3<Imag<16.3$ and $Bmag$ available from PPMXL. The stars are
selected to satisfy the requirements of the fiber positioning system and have a uniform distribution in
the $I$ vs. $B-I$ color-magnitude diagram. Our final input catalog consists of 12,000 objects on
each of 8 plates that are
observable during the winter observing season in Xinglong Station of
the National Astronomical Observatory of China.
 }

\authorrunning{L. Chen et al.}            
\titlerunning{The LEGUE disk targets for LAMOST's pilot survey}  

  \maketitle

%
%
\section{Introduction}           

\label{sect:intro} The Guoshoujing Telescope (GSJT, formerly called the
Large Area Multi-fiber Spectroscopic Telescope - LAMOST) is a 4
meter telescope with 4000 fibers in the focal plane (Cui et al. 2012). The telescope
is located at Xinglong Station, located 114 km to the east north
east (ENE) of Beijing and operated by the National Astronomical
Observatories of China. Currently, a pilot survey is running on the
telescope, in preparation for a full spectroscopic survey, which is
expected to begin in late 2012.  The survey has the potential to
significantly increase our understanding of the substructures in the
Galactic stellar spheroid and disk components through
measurements of radial velocities, metallicities and effective
temperatures for millions of stars (Deng et al. 2012).

Here we describe here a low latitude disk survey concentrating on open
clusters, which is being conducted as part of the LAMOST Pilot
Survey.  The pilot survey mainly aims to test the LAMOST system, in
preparation for the planned major survey. The pilot survey  targets
fainter objects on dark nights (Yang et al. 2012, Carlin et al.
2012), and brighter objects when the Moon is bright or the
atmosphere has low transparency (Zhang et al. 2012). The footprint
of the bright night disk survey is shown in Figure 1 overlaid on a
starcount map from SDSS photometry. The magenta part is the disk
survey region centered at b=$0^\circ$ which will be discussed in
this paper.

One of the primary science objectives of the LAMOST disk survey is
to investigate the structure of the thin/thick disks of the Galaxy,
including the chemical abundance as a function of position within
the disk and the extinction in the disk. In the main survey, we
expect to cover 300 open clusters in the low Galactic latitude
region, and obtain stellar radial velocities as well as abundance
information for stars as faint as $r=16^m$ in the cluster fields.
This will be the largest spectroscopic data sets for studying the
properties of Galactic open clusters, including the structure,
dynamics, and evolution of the disk as probed by open clusters (Chen
et al. 2003; Ahumada et al. 2011).

Some of the expected scientific outputs from the disk pilot survey include:
\begin{itemize}
\item significant improvements in essential parameter measurement of the targeted open clusters, using
kinematic membership information and homogeneous abundance data.
\item a database of spectroscopically confirmed cluster members in the outer parts of the clusters, that will provide good targets for detecting or verifying possible tidal tails of these stellar clusters.
\item a large sample of young stellar objects across the surveyed area, which will provide
important clues to properties of large-scale star formation and the history of Galactic
star formation as well as information on the 3D extinction in the Galactic plane.
\item tens of thousands of spectra of bright disk stars that can be used to characterize the chemistry and
kinematics of the thin disk.
\end{itemize}
The open clusters are particularly important to the survey, since some of the well characterized open
clusters provide important ``standards" for a range of stellar types that can be used to calibrate
the LAMOST observations and the data-processing pipeline.

In section 2, we describe the footprint of the survey as well as the catolog from which the targets
were selected.  In section 3, we describe the selection of spectroscopic targets, including selection
in color and magnitude, the effects of blending in a crowded field, and the properties of the selected
spectroscopic targets.
Finally, in section 4 we summarize the paper.


\section{Footprint and Input Catalog}
\label{sect:data}

For the disk pilot survey, we have chosen 8 plates along the Galactic plane that are nearly
uniformly distributed in Galactic longitude, and in a range of right ascension and declination
that will make them easily observable with GSJT.  Telescope design favors observations
with $-10^\circ<\delta<60^\circ$; lower declinations are unobservable, and the effective
aperture and image quality of the telescope decreases rapidly at declinations above $\delta=60^\circ$.
Since the observations should be taken near the meridian, it is
important to spread the target fields out in Galactic longitude so that there are targets available
at all right ascensions.  The part of the sky observed is $0^\circ<\alpha<67^\circ$, and
$42^\circ<\delta<59^\circ$.  Adjacent plates
may overlap slightly. Each plate covers a circle area of a radius $2.5^\circ$,
with central star (for the Shack-Hartmann guider) brighter than $V=8^m$.  It is a requirement
that each field observed by LAMOST have a bright star in the center so that the figure of the mirrors
can be actively adjusted to focus stellar images in the focal plane.

We were able to select fields that cover 22 previously identified star clusters.
Table 1 shows the selected open clusters within coverage of the designed disk pilot survey plates.
The angular distance between the cluster center and the plate center is less
than 2 degrees in all cases.  We also list here the parameters (Dias et al. 2002; WEBDA) for
each cluster, including (if available) cluster name, the position, reddening and distance, the
average proper motion and radial velocity, the cluster age estimation and the corresponding number
of the survey plate. Some of the clusters have been poorly studied and we expect to obtain
their kinematic and chemical information for the first time from the pilot survey database.


Individual stellar targets are selected from the PPMXL astrometric catalog (Roeser et al. 2010),
which includes positions, proper motions,  rough $BRI$ magnitudes, and $JHK$ magnitudes from 2MASS.
PPMXL is a full sky catalog of positions, proper motions, 2MASS and optical
photometry of 900 million stars and galaxies, aiming to be complete
for the brightest stars down to about $V=20^m$. It is the
result of a re-reduction of USNO-B1 observations, supplemented with 2MASS and ICRS
results. PPMXL is currently the largest collection of ICRS proper motions with typical
errors ranging from 4 mas yr$^{-1}$ to more than 10 mas yr$^{-1}$, depending
on the object's observational history.

In PPMXL, the photometric information from USNO-B1.0 is listed, as well as the
NIR photometry from 2MASS when available. The B,R,I magnitudes
in the USNO B1.0 magnitude system are rather crude, and there are
discrepancies in the magnitude system from field to field and from
early to late epochs. However, PPMXL is the only source that provides photometry in
optical bands for low galactic latitude targets, which  is
necessary as an input catalog for our disk pilot survey.

Figure 2 shows the star number density in one
of our plate regions, centered at ($\alpha=5.1421^\circ$,$\delta=56.5569^\circ$;
$l=118.6619^\circ$, $b=-6.0655^\circ$).  The PPMXL catlaog lists
$2.2\times 10^5$ point objects in this area.  On average there are over
11000 targets per square degree in this low Galactic latitude
region, complete in magnitude to I $\sim16.3^m$.

\begin{figure}
\centering
\includegraphics[width=10cm,angle=0]{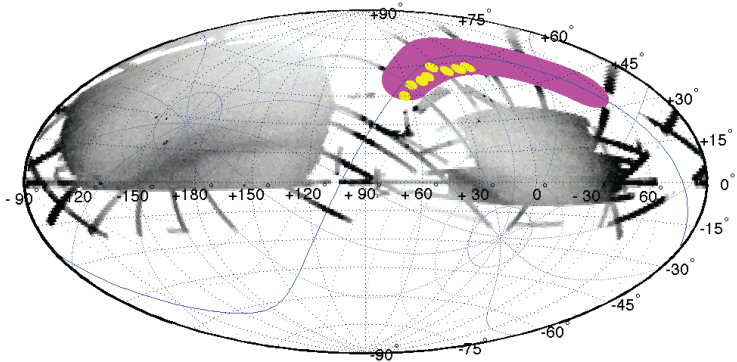}
\caption{The footprint of the bright night disk survey in equatorial coordinates
overlaid on a starcount map from SDSS photometry. The magenta disk survey region is centered
on the Galactic plane and the yellow circular areas are eight plates that were chosen for the pilot survey .  }
\label{Fig:footprint}
\end{figure}

\begin{figure}
\centering
\includegraphics[width=\textwidth,angle=0]{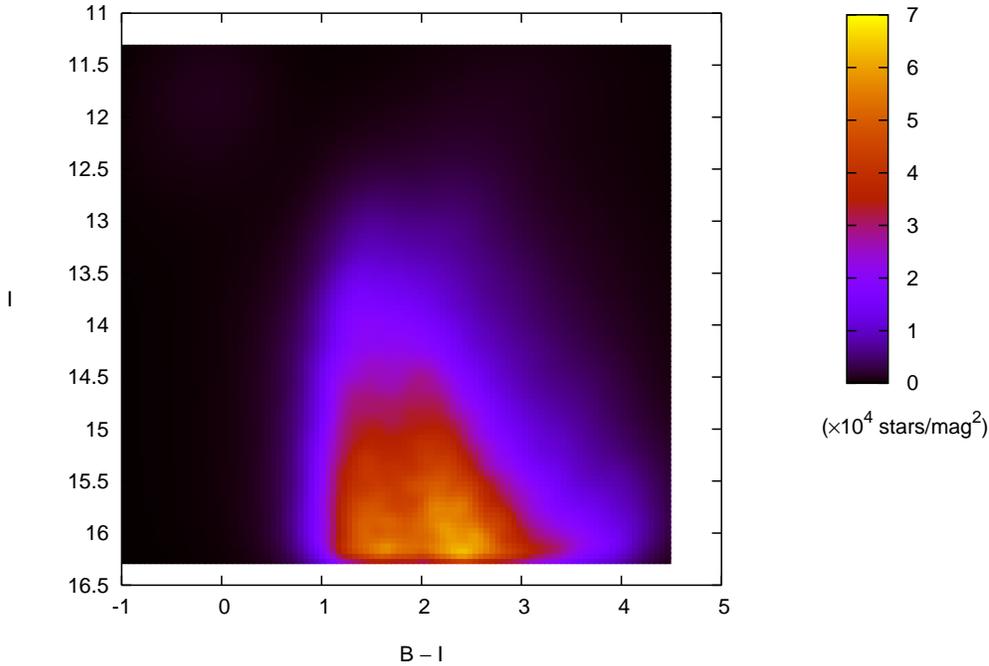}
\caption{The star number density of PPMXL in one of our plate regions, centered at $(RA=5.1421^\circ,DEC=56.5569^\circ)$.  A total of $2.2\times 10^5$ targets are included. The color bar on the right shows different number density scales in unit of $10^4 stars/mag^2$. }
\label{Fig:number-density}
\end{figure}

\section{Target Selection and Input Catalog Design}

\subsection{Target Selection Process}

For each of our selected plates, we must create an input catalog with about three times
as many stars as can be observed,
from which the fiber assignment software will select the actual targets.
Since GSJT
contains 200 fibers per square degree in the focal plane, we select
600 stars deg$^{-2}$ in the input catalog.
We explain here how we
select stars with $11.3< I_{\rm mag} <16.3$ and available $B_{\rm mag}$ from PPMXL.
Additionally, targets are selected to favor open cluster members, satisfy the requirements of the
fiber positioning system, and maintain a uniform distribution in I and B-I color-magnitude space.



Our target selection algorithm is as follows:

\begin{itemize}
\item For every LAMOST survey field,  we construct input catalogs of a bright subset of the PPMXL catalog with $11.3<I_{\rm mag}<14.3$ and a faint subset with $14.3<I_{\rm mag}<16.3$.

\item To prevent contamination of spectra by nearby stars (the blending effect, see discussion in the next section), we
eliminate all stars that have a similar brightness or a brighter neighbor
within 5". Note that the LAMOST fiber diameter is 3.3" and the typical seeing
in the Xinglong station is around 2" to 3".

\item To prevent local over density for fiber allocation, we set 6" as the minimum spatial
distance for all selected stars in the following steps.

\item We eliminate very high proper motion objects; the proper motion
limit is PM$<$100 mas yr$^{-1}$. Note that the typical mean observation epochs for PPMXL sources were during the 1970s or 1980s.

\item Highest priority is assigned to cluster members, whenever membership
information is available in Kharchenko's catalog (Kharchenko et al. 2005). Membership in this catalog is estimated from the proper motion of individual stars in the cluster region.

\item For the remainder of the targets, we perform uniformly sampling in
color-magnitude space (B-I vs.I) by randomly selecting stars on the color-magnitude
diagram (CMD).  For instance, for the bright subset of Plate 1 we set the magnitude range as $11.3^m <I< 14.3^m$ and color range as $-1.5^m <(B-I)< 4.5^m$ which
includes a total of about 36000 stars. First we generate a random point in the $(B-I)\sim I$ space, then within a search radius of 0.03 mag (roughly the average distance in the color-magnitude space) select the object closest to the random point as our sample star. In addition, all the selected stars will satisfy the spatial distance limit of 6". The iteration ends up with a bright sub-catalog of 12000 stars ready for the fiber assignment software to select the actual targets.  Here we note that the  magnitude from PPMXL is rather crude.

\end{itemize}

It was intended that the bright and faint datasets would be sent to fiber assignment separately, and two
pointings (one brighter and one fainter) would be done at each position.  Instead, the two lists
were merged before fiber assignment for the pilot survey, and in some cases a second or third
pointing was constructed (with many repeated sources).


\subsection{Blending in a crowded field}
\begin{figure}[!htpb]
\begin{center}
\includegraphics[scale=0.8]{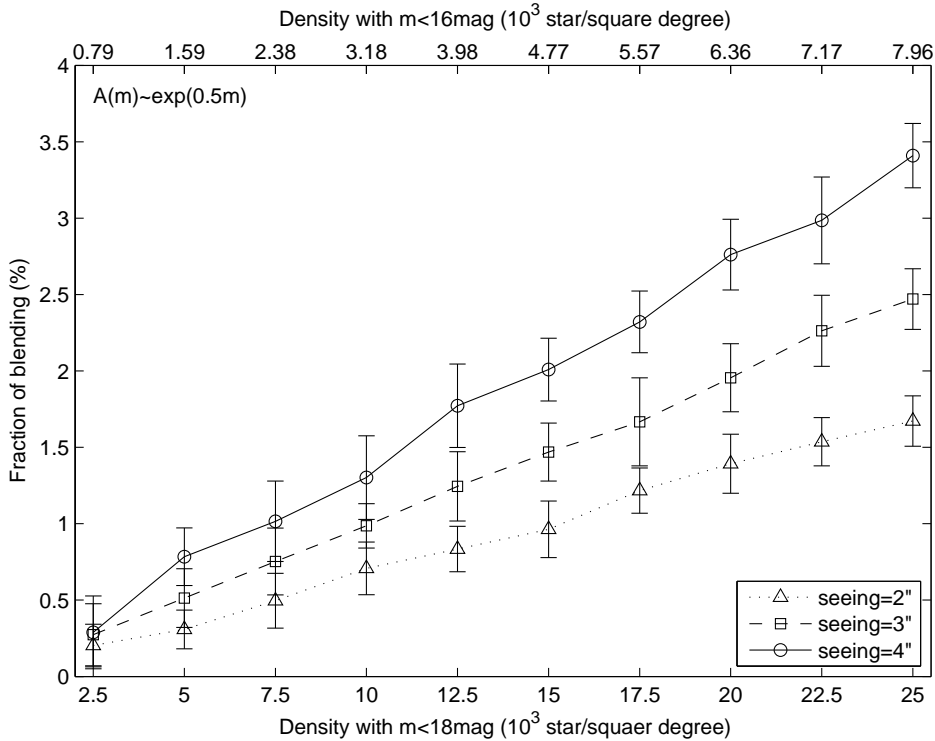}
\caption{The simulation of the blending within a fiber aperture. The X-axis on bottom is the density of the simulated objects with m$<18$\,mag, while on top is the density of the objects with m$<16$\,mag corresponding to the bottom X-axis. The luminosity function is assumed to be $A(m)\sim exp(0.5m)$, which is approximately estimated from SDSS photometry. The dotted, dashed, solid lines with triangles, rectangles and circles show the probability of more than 10\% of the flux of the blended objects being in the fiber aperture when fibers are assigned to all objects with m$<16$\,mag given the dome seeing being 2, 3, and 4 arcsec, respectively. The error is estimated from a Monte Carlo approach.}\label{fig:fiberblend}
\end{center}
\end{figure}

For a fiber fed multiobject spectroscopic telescope such as GSJT, blending, which can be defined as
at least two objects are captured by a fiber in an exposure,
must be taken into account when the telescope targets crowded fields, (e.g.
the Galactic disk, open clusters).

LAMOST uses 4000 fibers on the 1.75-meter-diameter focal plane; each fiber
has an aperture of 3.3\,arcsec. When a fiber is targeting an object, some flux contributed by its neighboring
objects may also be recorded. Hence, the final processed
spectra could be very strange if it is contaminated by the blended
objects. In order to investigate the probability of the blending in
LEGUE disk survey, we make a simple simulation.

We assume that 1) all objects in the simulation are point sources; 2) the range of the brightness of the objects covers $10<m<18$\,mag, the fainter objects are ignored; 3) we only ``observe" objects brighter than 16\,mag, the objects between 16 and 18\,mag only contribute to the blending; 4) the luminosity function is studied using SDSS photometry catalog and it can be approximately fitted by $A(m)\sim exp(0.5m)$; and 5) the PSF is a Gaussian, in which the $\sigma$ is smaller than the dome seeing by a factor of 2.35.

In the simulation we consider a circular, one square degree region filled with uniformly
distributed objects (down to 18\,mag). The density is from 2,500 to 25,000 stars/square degree,
which is the typical density range in the outer disk and open clusters. Then we assign fibers for
all objects with $m<16$\,mag given the dome seeing. For each fiber we calculate the ratio of the
total flux from the neighboring objects (only those located within 3$\times$seeing are considered)
to that from the target in the fiber aperture. If the ratio is higher than 10\%, we mark this fiber
as a significantly blended fiber. Note that the sky flux is ignored in the calculation. The fraction
of the significantly blended fibers in the field is defined as the probability of the blending.

We did the simulation for the seeing of 2", 3" and 4". The results are shown in
Figure 3. We find that the probability of more than 10\%of the flux of the blended
objects being in the fiber aperture increases linearly with the density of the objects. Moreover,
the bigger the seeing the higher the probability of the blending.  For instance, in the center region
of an open cluster, in which the density can be around 10,000 per square degree, the LAMOST
spectra will suffer significant blending in about 0.7\% of the spectra on each plate, given a seeing of 2".
It increases to 1\% and 1.3\% when the seeing is increaseed to 3" and 4", respectively. It is worthy to note that in practice
even fainter neighboring objects, e.g. objects with strong emission lines, out skirt of a neighboring
galaxy, nearby emission nebulae etc., may also contribute to the flux. This limit will constrain
the target selection for the disk survey to either avoid very crowded regions or accept the
fact that there would be a few percent blended spectra and later develop a pipeline to address this issue.

\subsection{Sample Plate for the Disk Portion of the Pilot Survey}

As an example of the outputs from the target selection process,
Figure 4 shows the color-magnitude distribution of
selected targets as a bright subset (left panel) $(11.5<I_{\rm mag}<14.3)$
for Plate 1 and the spatial distribution (right panel) of selected
stars. It can be seen that our uniform selection in
color-magnitude space favors bluer as well as brighter
objects in the input catalog. There are two open clusters well
within the field of view of Plate 1 (centered at  RA=5.1421$^\circ$,
DEC=56.5569$^\circ$) : ASCC 2 and ASCC 3. We assign highest priority to
members (87 stars) of these clusters (membership information from
Kharchenko et al. 2005). Here the output of our selection algorithm
largely achieved the basic goal for target selection from field disk stars:
we would like to sample a larger fraction of the rarer stars than the less rare,
and a larger fraction of the 진interesting진 than the 진less
interesting.진

In this plate, we expect to observe all 87 open cluster member
stars which can be used to calibrate the spectra quality and
stellar parameters, especially the radial velocity and metallicity.

Figure 5 shows the color-magnitude distribution of targets in a faint subset (left panel) $(14.3<I_{\rm mag}<16.3)$
and the corresponding spatial distribution (right panel) of selected stars. Since the magnitude limit in Kharchenko's catalog (Kharchenko et al. 2005)
is brighter than $V\sim 14^m$, there are no membership information available for this faint subset.
In the future, we will use astrometric
catalogs (e.g. PPMXL and UCAC3) to determine cluster membership to a deeper magnitude limit from proper
motion.

\begin{figure}
\centering
\includegraphics[width=\textwidth,angle=0]{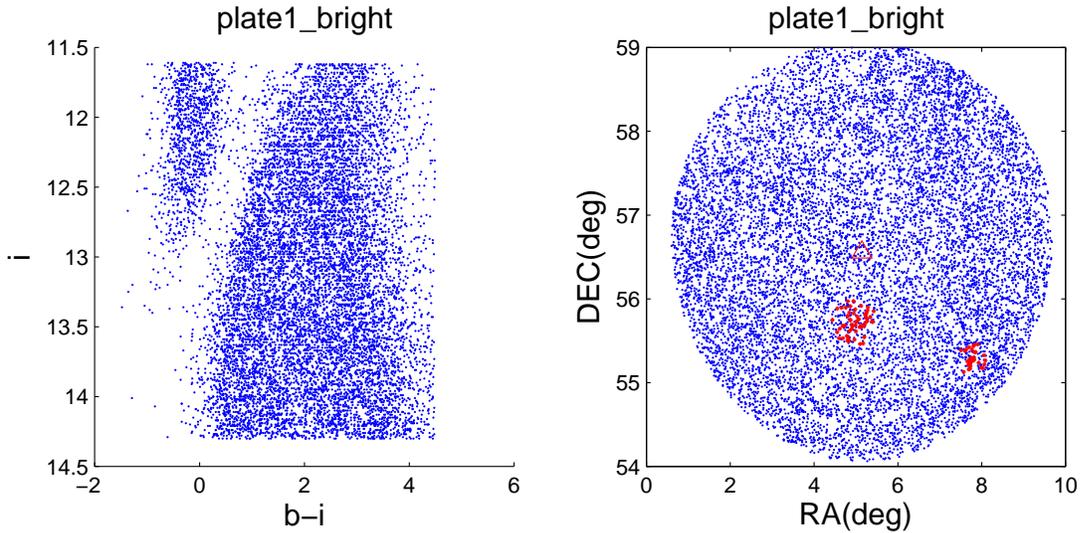}
\caption{Left: the color-magnitude distribution of selected targets as a bright subset
$(11.3<I_{\rm mag}<14.3)$ for plate1, satisfying requirement for SSS fiber-assignment;
right:spatial distribution of selected stars. Red *: OC members, red triangle: plate center} \label{Fig:Plate-1}
\end{figure}

\begin{figure}
\centering
\includegraphics[width=\textwidth,angle=0]{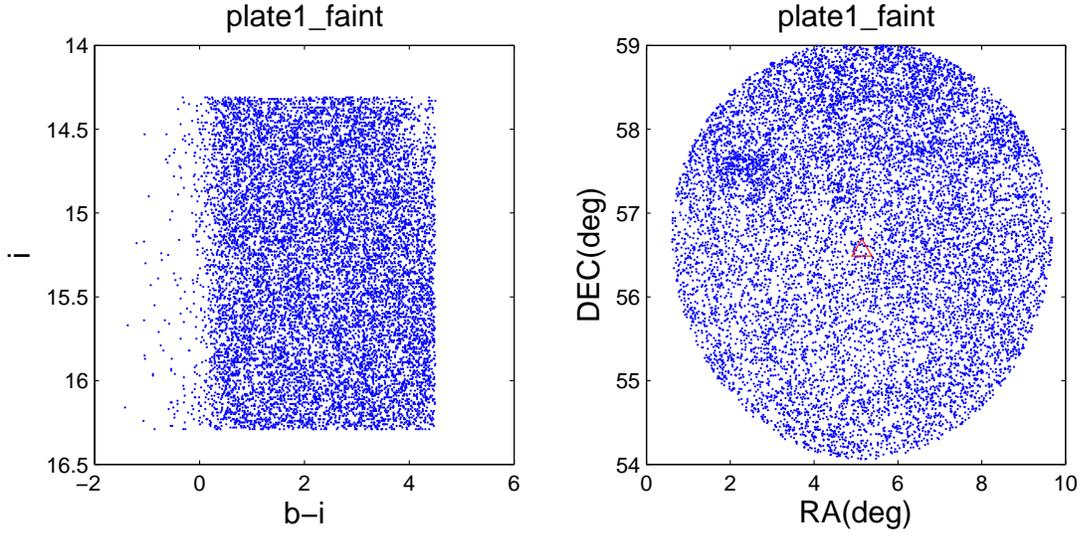}
\caption{Left: the color-magnitude distribution of selected targets
as a faint subset $(14.3<I_{\rm mag}<16.3)$ for Plate-1, satisfying
requirement for SSS fiber-assignment; right:spatial distribution of
selected stars. Red triangle: plate center}
\end{figure}

\section{Summary}
We have described the target selection algorithm for eight low latitude, bright plates that are
planned to be observed during the LAMOST Pilot Survey.  Stars from 22 open clusters will be observed
on these plates.
To date, 5 plates have been observed successfully, and the data reductions and quality assessment
is being conducted.  This portion of the Pilot Survey is in preparation for a larger survey that
will cover 300 Galactic open clusters.

The highest priority targets on these plates are known or suspected members of Galactic open clusters.
Other disk stars in the $11.3<I_{\rm mag}<16.3$ range were selected uniformly over I, B-I color-magnitude
phase space.  Very high proper motion objects are not targeted.  Positions, proper motions, and
magnitudes are taken from the PPMXL catalog.

The spectra observed in this disk portion of the LAMOST Pilot Survey will be used for studies
of Galactic open clusters, to investigate young stellar objects, and to study the chemistry
and kinematics of disk stars.  The open clusters stars will be important calibrators for the main
spectroscopic survey.

\begin{table}[!t]
\caption{Properties of open clusters in the disk pilot survey area.}
\begin{center}
\begin{tabular}{|c|c|c|c|c|c|c|c|c|c|c|c|}

\hline

Name & RA    & Dec   & l          & b          & Ebv & Dist & $\mu_{\alpha} cos\delta$ & $\mu_{\delta}$ & Rv   & Age & Plate Nr. \\
    & h m s & d m s & ($^\circ$) & ($^\circ$) &     & pc   &        mas/yr            &         mas/yr & km/s & Gyr &           \\
\hline

ASCC 2      & 0 19 51.6 & 55 42 36  & 118.46  & -6.89 &  0.10 &  1200 &  -0.91  & -3.94 &        & 0.676 & 1 \\
Stock 21    & 0 29 49.2 & 57 55 12  & 120.05  & -4.83 &  0.40 &  1100 &  -0.12  & -1.52 &        & 0.525 & 1 \\
ASCC 3      & 0 31  8.4 & 55 16 48  & 120.02  & -7.48 &  0.17 &  1700 &  -1.92  & -1.25 & -37.00 & 0.079 & 1 \\
King 2      & 0 50 60.0 & 58 10 48  & 122.87  & -4.69 &  0.31 &  5750 &         &       &        & 5.023 & 2 \\
IC 1590     & 0 52 48.0 & 56 37 48  & 123.12  & -6.24 &  0.32 &  2384 &  -1.36  & -1.34 & -37.50 & 0.007 & 2 \\
ASCC 5      & 0 57 57.6 & 55 50 24  & 123.85  & -7.02 &  0.25 &  1500 &  -3.10  & -2.91 & -43.00 & 0.011 & 2 \\
NGC 657     & 1 43 20.9 & 55 50 24  & 130.22  & -6.30 &       &       &         &       &        &       & 3 \\
ASCC 6      & 1 47 13.2 & 57 43 48  & 130.34  & -4.34 &  0.30 &  1200 &  -1.02  & -1.18 & -20.00 & 0.148 & 3 \\
NGC 743     & 1 58 37.0 & 60  9 36  & 131.20  & -1.63 &       &       &         &       &        &       & 4 \\
ASCC 7      & 1 58 55.2 & 58 58 12  & 131.54  & -2.77 &  0.50 &  2000 &  -0.57  & -3.08 & -49.00 & 0.023 & 4 \\
Stock 2     & 2 14 24.0 & 59 16 12  & 133.36  & -1.92 &  0.34 &   380 &  16.22  &-13.39 &  -2.85 & 0.148 & 4 \\
NGC 869     & 2 19  4.8 & 57  8 60  & 134.63  & -3.72 &  0.48 &  2079 &  -0.49  & -0.90 & -39.82 & 0.019 & 4 \\
Basel 10    & 2 19 28.1 & 58 17 60  & 134.30  & -2.62 &  0.77 &  1944 &   1.47  &  0.35 &        & 0.041 & 4 \\
ASCC 8      & 2 20 49.2 & 59 36 36  & 134.02  & -1.33 &  0.55 &  2200 &  -1.24  &  0.57 & -42.09 & 0.006 & 4 \\
NGC 884     & 2 22  1.2 & 57  8 24  & 135.01  & -3.60 &  0.56 &  2345 &  -0.84  & -0.23 & -38.14 & 0.013 & 4 \\
Trumpler 2  & 2 36 54.0 & 55 55 12  & 137.37  & -3.97 &  0.32 &   649 &   1.40  & -5.57 & -39.00 & 0.085 & 5 \\
Czernik 12  & 2 39 12.0 & 54 55 12  & 138.08  & -4.75 &       &       &         &       &        &       & 5 \\
NGC 1220    & 3 11 40.1 & 53 20 24  & 143.04  & -3.96 &  0.70 &  1800 &         &       &        & 0.060 & 6 \\
King 5      & 3 14 36.0 & 52 43 12  & 143.74  & -4.27 &       &  1900 &         &       & -52.00 & 0.759 & 6 \\
Czernik 15  & 3 23 12.0 & 52 15  0  & 145.10  & -3.97 &       &       &         &       &        &       & 6 \\
King 7      & 3 59  0.0 & 51 47 60  & 149.77  & -1.02 &  1.25 &  2200 &         &       &        & 0.661 & 7 \\
Berkeley 11 & 4 20 36.0 & 44 55 12  & 157.08  & -3.64 &  0.95 &  2200 &   1.45  & -2.37 &        & 0.110 & 8 \\

\hline
\end{tabular}
\end{center}
\label{tab:frac_select}
\end{table}%

\begin{acknowledgements}
We thank the referee, Sanjib Sharma, for helpful comments on the manuscript.
This work was funded by the National Natural Science Foundation of
China (Grant Nos. 11173044(PI:Hou),11073038(PI:Chen), 10573022, 10973015,11061120454
(PI:Deng), the Key Project No.10833005(PI:Hou),the Group Innovation
Project No.11121062, and the US National Science Foundation (grant AST 09-37523. Chinese Academy of Science is acknowledged for providing  initial support from grant number GJHZ200812.
\end{acknowledgements}

\appendix                  

\label{lastpage}

\end{document}